\documentclass[aip,pop,preprint,a4paper]{revtex4-1}
\usepackage{cancel}
\usepackage{amsmath}
\usepackage{amssymb}
\usepackage{graphicx}% Include figure files
\usepackage{dcolumn}% Align table columns on decimal point
\usepackage{bm}% bold math
\usepackage{upgreek}

\draft % marks overfull lines with a black rule on the right

\begin{document}

%\preprint{}

\title{Magnetohydrodynamic motion of a two-fluid plasma} %Title of paper



\author{J. W. Burby}
 \affiliation{Courant Institute of Mathematical Sciences, New York, New York 10012, USA}
%\thanks{}


\date{\today}

\begin{abstract}
The two-fluid Maxwell system couples frictionless electron and ion fluids via Maxwell's equations. When the frequencies of light waves, Langmuir waves, and single-particle cyclotron motion are scaled to be asymptotically large, the two-fluid Maxwell system becomes a fast-slow dynamical system. 
This fast-slow system admits a formally-exact single-fluid closure that may be computed systematically with any desired order of accuracy through the use of a functional partial differential equation. In the leading order approximation, the closure reproduces magnetohydrodynamics (MHD). Higher order truncations of the closure give an infinite hierarchy of extended MHD models that allow for arbitrary mass ratio, as well as perturbative deviations from charge neutrality.
The closure is interpreted geometrically as an invariant slow manifold in the infinite-dimensional two-fluid phase space, on which two-fluid motions are free of high-frequency oscillations. This perspective shows that the full closure inherits a Hamiltonian structure from two-fluid theory. By employing infinite-dimensional Lie transforms, the Poisson bracket for the all-orders closure may be obtained in closed form. Thus, conservative truncations of the single-fluid closure may be obtained by simply truncating the single-fluid Hamiltonian. Moreover, the closed-form expression for the all-orders bracket gives explicit expressions for a number of the full closure's conservation laws. Notably, the full closure, as well as any of its Hamiltonian truncations, admits a pair of independent circulation invariants.
\end{abstract}


\maketitle %\maketitle must follow title, authors, abstract and \pacs


\section{Introduction}
Ideal magnetohydrodynamics\cite{Friedberg_2014} (MHD) is a well-known reduced plasma model that treats a plasma as a single conducting fluid. Because real plasma is made up of a large collection of discrete particles, it is natural to wonder how such a single-fluid model could have any predictive capability. This challenging problem has been addressed on numerous occasions\cite{Kulsrud_2005}, and most assessments conclude that MHD does a good job of predicting plasma equilibrium and stability at ``large-scales.'' However, this answer is unsatisfactory for a variety of reasons. Most importantly, it does not tell us \emph{why} MHD works, only that it does. In contrast, there must be some physical mechanism that enables a many-particle-field system to exhibit MHD behavior. Previous investigations provide only vague suggestions of what this mechanism might be.

The purpose of this article is to study an important part of this mechanism. Before elaborating further, however, the sense in which a ``part'' of the mechanism can even be discussed is worthy of explanation. One possible way to exhibit a mechanism for emergent MHD behavior in a many-particle system is to first show and explain emergent multi-fluid behavior, and then explain how MHD emerges from multi-fluid dynamics. This approach naturally breaks the mechanism into two parts, and this article will discuss (the simpler) one, namely the submechanism by which MHD behavior emerges from the dynamics of multiple charged fluids. Of course, if there is no submechanism for multi-fluid dynamics to emerge in a many-particle model, then the discussion contained in this article would be neither novel nor useful. Therefore the assumption that multi-fluid dynamics can indeed be found within many-particle dynamics will be tacitly assumed henceforth.

Roughly speaking, the dynamical content that is missing from MHD consists of rapidly oscillating modes, including Langmuir waves and light waves. Therefore one tempting explanation for the emergence of MHD motion in a two-fluid plasma is the effective damping of these rapidly varying modes. Even though the ideal two-fluid-Maxwell system does not include collisional dissipation, this damping mechanism may still be present as a result of phase mixing\cite{Bedrossian_2013}. A second possible explanation may be that the rapidly oscillating modes do not damp, but instead are effectively averaged out. If this explanation is valid, it would be especially interesting because it would suggest that there must be some kind of ponderomotive forcing\cite{Cary_pond_1981} of the MHD state variables by the rapidly oscillating modes that has not been calculated previously. Finally, there is at least one other possible explanation. Perhaps there are special initial configurations of a two-fluid plasma that do not excite the rapidly oscillating modes at all. In other words, it may be that Langmuir waves and light waves are neither damped, nor averaged-out, but instead fail to be excited in the first place. While all of these possible explanations for emergent MHD behavior may be interesting, interrelated, or perhaps mutually independent, the ensuing discussion and analysis will focus on the third possible mechanism, which is convenient to refer to as ``lazy high-frequency modes."

An oversimple caricature of lazy high-frequency modes consists of two pendula placed in a room, one much longer than the other. The most general motion of these pendula (assuming small amplitude oscillations, for simplicity) involves each pendulum swaying at its respective characteristic frequency, and therefore involves a pair of disparate time scales. However, there are special ``slow'' motions of the system wherein the short pendulum is motionless, meaning that the fast time scale in the problem is not present. Note that these motions are characterized by special initial conditions that allow only the long pendulum to be displaced from its equilibrium location in phase space.  Also note that the slow dynamics is governed by Newton's Second Law applied to just the long pendulum, which is a dynamical system whose dimension is less than that of the total system. Here the short pendulum is the analogue of the rapidly-varying modes in the two-fluid model, while the long pendulum represents MHD motion. Although the special ``slow'' initial conditions are obvious in this toy problem, the same cannot be said of two-fluid dynamics. There, all of the modes are coupled nonlinearly, and so it is not clear that slow initial conditions even exist, let alone possess a simple parameterization.

In order to argue that slow initial conditions for two-fluid dynamics \emph{do} exist, this article will deduce three technical results: (a) an asymptotic expansion for the set of slow two-fluid initial conditions, (b) an asymptotic expansion of the reduced dynamical equations that govern slow dynamics, and (c) the variational and Hamiltonian structures underlying the slow dynamics, which are naturally inherited from the corresponding structures underlying two-fluid dynamics. Interestingly, the mathematical tools that will lead to these results are powerful enough to provide a simple, closed-form expression for the slow dynamics' Poisson bracket. 

Modulo delicate issues related to convergence of the asymptotic expansions (see Section \ref{discussion} for a discussion of this point), these results will show: (1) that there is a collection of slow initial conditions for the two-fluid system of equations that is parameterized by the MHD phase space, (2) that the reduced equations governing the slow dynamics are equivalent to an extended MHD model with low-order truncations that reproduce ideal MHD and Hall MHD, and (3) that the Hall MHD Poisson bracket governs the slow dynamics to all orders. Taken together, (1) and (2) imply that lazy high-frequency modes may indeed be a plausible mechanism for MHD-like motion of a two-fluid plasma. Moreover, (3) implies that the problem of developing dissipation-free approximations of this MHD-like motion reduces to finding an approximate expression for the slow dynamics' Hamiltonian functional. This is a desireable feature for an extended MHD theory to have, for instance, when using such a model to study collisionless reconnection.

\section{Two-fluid dynamics: scaling and variational principle}
The asymptotically-scaled ideal two-fluid-Maxwell system is given by
\begin{gather}
\hspace{-4em}m_i n_i (\partial_t\bm{u}_i+\bm{u}_i\cdot\nabla\bm{u}_i)=-\nabla \mathsf{p}_i\label{ion_momentum}\\
\hspace{8em}-\frac{1}{\epsilon}Z_iq_e n (\bm{E}+\bm{u}_i\times\bm{B})\nonumber\\
\hspace{-4em}m_e(Z_i n_i +\epsilon^2 \epsilon_o q_e^{-1}\nabla\cdot\bm{E})(\partial_t\bm{u}_e+\bm{u}_e\cdot\nabla\bm{u}_e)=\label{electron_momentum}\\
\hspace{1em}-\nabla\mathsf{p}_e+\frac{1}{\epsilon}q_e (Z_i n_i +\epsilon^2 \epsilon_o q_e^{-1}\nabla\cdot\bm{E})(\bm{E}+\bm{u}_e\times\bm{B})\nonumber\\
\partial_t n_i+\nabla\cdot(n_i \bm{u}_i)=0\label{ion_continuity}\\
\nabla\times\bm{B}=\frac{1}{\epsilon}\mu_oq_eZ_i n_i(\bm{u}_e-\bm{u}_i)\label{ampere}\\
\hspace{8em}+\epsilon\,\mu_o\epsilon_o(\partial_t\bm{E}+\bm{u}_e\nabla\cdot\bm{E})\nonumber\\
\nabla\times\bm{E}=-\partial_t\bm{B}\label{faraday}
\end{gather}
where $n_i$ is the ion number density, $\bm{u}_\sigma$ is the velocity of species $\sigma$, $\mathsf{p}_\sigma$ is the species-$\sigma$ partial pressure, $\bm{B}$ is the divergence-free magnetic field, $\bm{E}$ is the electric field, $m_\sigma$ is the species-$\sigma$ mass, $Z_i$ is the ionic atomic number, $q_e$ is (minus) the elementary unit of charge, and $\mu_o,\epsilon_o$ are the usual MKS vacuum permeability and permittivity. I will assume that the spatial domain is the $3$-torus; non-periodic boundary conditions will require a separate analysis. Upon setting the mass ratio $m_e/m_i=\nu$, it is also useful to introduce the scalar fields
\begin{gather}\label{scalars}
v_A^2=\frac{1}{1+\nu Z_i}\frac{\mu_o^{-1} |\bm{B}|^2}{m_i n_i}\quad \omega_p^2= (1+\nu Z_i)\frac{q_e^2 Z_i n_i}{\epsilon_o m_e}\\
\omega_{c i}=-\frac{q_e Z_i |\bm{B}|}{m_i},\nonumber
\end{gather}
which represent the (squared) speed of Alfv\'en waves, the (squared) frequency of Langmuir oscillations, and the frequency of ion cyclotron motion.

In their order of appearance, the equations comprising the two-fluid Maxwell system express the conservation of momentum for ions and electrons, the conservation of ion number,  the Amp\'ere-Maxwell Law, and Faraday's Law. For the sake of simplicity, I have assumed a barotropic equation of state for both electrons and ions
\begin{align}
\mathsf{p}_i&=\mathsf{p}_i(n_i)\\
\mathsf{p}_e&=\mathsf{p}_e(Z_i n_i +\epsilon^2 \epsilon_o q_e^{-1}\nabla\cdot\bm{E}),
\end{align}
but most of the ensuing discussion would only be modified superficially upon adopting an equation of state that accounts for entropic dynamics. Notice that the electron number density does not appear explicitly in the two-fluid Maxwell system as it is written above. This is accomplished by using Gauss' Law to eliminate the electron number density in favor of the electric field and the ion number density. The two-fluid state is therefore given by the tuple of fields $Z=(n_i,\bm{u}_i,\bm{B},\bm{u}_e,\bm{E})$, and the two-fluid-Maxwell system may be regarded as a first-order ODE on $Z$-space, i.e. $Z$-space is the (infinite-dimensional) two-fluid-Maxwell phase space. Note that if the electron number density were not eliminated, the two-fluid Maxwell system would instead take the form of a differential-algebraic system on a slightly larger space. Because I will evetually use some ideas from dynamical systems theory to analyze two-fluid dynamics, this would be a technical inconvenience.

The unscaled ($\epsilon=1$) two-fluid-Maxwell system, which is perhaps more familiar than the scaled version, may be transformed into the scaled two-fluid-Maxwell system by making the simple substitutions
\begin{align}
q_e\mapsto \frac{1}{\epsilon}q_e\\
\epsilon_o\mapsto\epsilon\epsilon_o.
\end{align}
Formally, rescaling the elementary charge and vacuum permittivity is equivalent to working with dimensionless variables and then adopting the ``drift-kinetic ordering''
\begin{gather}
\frac{\omega}{\omega_{ci}}\sim\frac{\rho_i}{L}\sim \frac{v_{the}^2}{c^2}\sim\epsilon\\
\beta\sim\frac{m_e}{m_i}\sim\frac{u_o}{v_{the}}\sim\frac{u_E}{v_{the}} \sim 1,
\end{gather}
where $v_{the}$ is the electron thermal velocity, $\rho_i=\sqrt{m_e/m_i}v_{th e}/\omega_{ci}$ is the ion gyroradius, $c=\sqrt{\mu_o\epsilon_o}^{-1}$ is the speed of light, $\beta=\mu_o m_i n_i v_{th i}^2/B_o^2$ is the plasma $\beta$, $u_o$ is the characteristic flow speed, $u_E=E_o/B_o$ is the characteristic $E\times B$ speed, and $\omega,L$ represent the characteristic time and length scales for the slow dynamics. In this article, the dimensional MKS unit system will be adopted instead of the natural dimensionless unit system implied by the drift kinetic ordering. Physical expressions with the correct units may therefore always be recovered by setting $\epsilon=1$. Physically, the drift kinetic ordering implies that the observation time scale is much longer than the ion cyclotron period, while the observation length scale is much longer than the ion gyroradius. Thus, the usual assumptions underlying guiding center theory are valid in the drift kinetic ordering. The physical interpretation of the remainder of the drift kinetic ordering is: non-relativistic electrons ($\frac{v_{the}^2}{c^2}\sim\epsilon$), arbitrary mass ratio and plasma beta ($\beta\sim\frac{m_e}{m_i}\sim 1$), bulk flow speed comparable to $E\times B$-speed, and $E\times B$-speed comparable to the electron thermal speed ($\frac{u_o}{v_{the}}\sim\frac{u_E}{v_{the}} \sim 1$). This scaling will be leveraged in what follows to find asymptotic expansions for both the special slow two-fluid configurations as well as the reduced evolution law for the slow dynamics. 

An important property of the scaled two-fluid-Maxwell system is that it may be derived from a phase space variational principle. The phase space Lagrangian is given by
\begin{align}\label{phase_space_lagrangian}
L&=\int (Z_i n_i +\epsilon^2 \epsilon_o q_e^{-1}\nabla\cdot\bm{E})(m_e \bm{u}_e+\epsilon^{-1}q_e \bm{A})\cdot\bm{v}_e\,d^3\bm{x}\nonumber\\
&+\int n_i (m_i\bm{u}_i-\epsilon^{-1}Z_i q_e \bm{A})\cdot\bm{v}_i\,d^3\bm{x}\nonumber\\
&-\int\epsilon \epsilon_o\bm{E}\cdot\dot{\bm{A}}\,d^3\bm{x}-\mathcal{H},
\end{align}
where the Hamiltonian functional is given by
\begin{align}
\mathcal{H}&=\frac{1}{2}\int m_i n_i |\bm{u}_i|^2\,d^3\bm{x}+\int n_i \,\mathcal{U}_i(n_i)\,d^3\bm{x}\nonumber\\
&+\frac{1}{2}\int m_e (Z_i n_i +\epsilon^2 \epsilon_o q_e^{-1}\nabla\cdot\bm{E})|\bm{u}_e|^2\,d^3\bm{x}\nonumber\\
&+\int (Z_i n_i +\epsilon^2 \epsilon_o q_e^{-1}\nabla\cdot\bm{E})\,\mathcal{U}_e(Z_i n_i +\epsilon^2 \epsilon_o q_e^{-1}\nabla\cdot\bm{E})\,d^3\bm{x}\nonumber\\
&+\frac{1}{2}\int\epsilon \epsilon_o|\bm{E}|^2\,d^3\bm{x}+\frac{1}{2}\int \mu_o^{-1}|\bm{B}|^2\,d^3\bm{x}.
\end{align}
The functions $\mathcal{U}_e(n_e)$ and $\mathcal{U}_i(n_i)$ are the internal energy densities for electrons and ions respectively. They are determined up to unimportant additive constants by the thermodynamic identities
\begin{align}
\mathsf{p}_\sigma(n_\sigma)=n_\sigma^2\,\mathcal{U}_\sigma^\prime(n_\sigma).
\end{align} 
When computing the Euler-Lagrange equations associated with $L$, variations are taken with respect to $(n_i,\bm{u}_i,\bm{A},\bm{u}_e,\bm{E},\bm{v}_i,\bm{v}_e)$. The fields $(\bm{u}_i,\bm{u}_e,\bm{A},\bm{E})$ are subject to arbitrary variations, while the fields $(n_i,\bm{v}_i,\bm{v}_e)$ are subject to constrained variations, as is standard in Euler-Poincar\'e variational principles\cite{Holm_1998},
\begin{align}
\delta n_i&=-\nabla\cdot(n_i\bm{\xi}_i)\\
\delta \bm{v}_i&=\dot{\bm{\xi}}_i+\bm{v}_i\cdot\nabla \bm{\xi}_i-\bm{\xi}_i\cdot\nabla\bm{v}_i\\
\delta \bm{v}_e&=\dot{\bm{\xi}}_e+\bm{v}_e\cdot\nabla \bm{\xi}_e-\bm{\xi}_e\cdot\nabla\bm{v}_e.
\end{align}
Here $\bm{\xi}_{i,e}$ are arbitrary vector fields defined on the fluid domain. They represent Eulerian displacements of the ion and electron fluids that are generated by variations of the Lagrangian trajectories. While Ref.\,\onlinecite{Holm_1998} goes into much more detail on this point, it is worth mentioning that the constrained variations of $(n_i,\bm{v}_i,\bm{v}_e)$ are actually consequences of unconstrained variations of the Lagrangian coordinates of electron and ion fluid parcels.

\section{\label{invariance_equations}Parameterization and time evolution of slow initial conditions}
Define a \emph{slow initial configuration} of a two-fluid plasma as an initial condition for the scaled two-fluid-Maxwell system that is $O(1)$ and whose time derivative is $O(1)$. The condition on the time derivative is reasonable because the drift-kinetic ordering involves equating the dynamical time scale with a time scale that is long when compared with the cyclotron period. The question this section will address is ``what are the slow initial configurations, and how do they evolve in time?" Satisfying answers will be obtained in an asymptotic sense.

Consider first two basic mathematical properties of the slow initial conditions. 
\\ \\
\noindent \emph{property 1:} If $Z$ is a slow initial state, then the state $Z(T)$ obtained by letting $Z$ evolve for $ T$ seconds according to the two-fluid-Maxwell system is also a slow initial state, regardless of the value of $T$. To see this, first observe that since the path $Z(t)$ is free of rapid oscillations, so too is the path $Z^\prime(t)=Z(t+T)$. But, by the uniqueness of solutions to ordinary differential equations (e.g. the two-fluid-Maxwell system), $Z^\prime(t)$ is precisely the solution of the two-fluid-Maxwell system with initial condition $Z(T)$. In other words, $Z(T)$, regarded as an initial condition, produces a slow evolution. This proves the claim.
\\ \\
\noindent \emph{remark:} Property 1 is a statement pertaining to the whole set of slow initial conditions. In the context of dynamical systems theory\cite{Wiggins_2003}, the collection of slow initial conditions would be referred to as an \emph{invariant set}. More generally, given a dynamical system on some phase space, an invariant set is a subset of phase space with the property that points in the subset stay inside the subset under dynamical evolution.
\\ \\
\noindent \emph{property 2:} As $\epsilon\rightarrow 0$, the set of slow initial conditions, $S_\epsilon$ (which is a subset of $Z$-space), must contain
\begin{align}
S_0=\{(n_i,\bm{u}_i,\bm{B},\bm{u}_e,\bm{E})\mid \bm{u}_e=\bm{u}_i,\hspace{.5em} \bm{E}=-\bm{u}_i\times\bm{B}\}.
\end{align}
In order to verify that this is true, suppose that $Z_\epsilon=(n_{i\epsilon},\bm{u}_{i\epsilon},\bm{B}_\epsilon,\bm{u}_{e\epsilon},\bm{E}_\epsilon)$ is a slow solution of the scaled two-fluid-Maxwell system. Because the solution is slow, the time derivative of $Z_\epsilon$ must be $O(1)$. Therefore the terms in Eqs.\,(\ref{ion_momentum})-(\ref{faraday}) that are multiplied by $\epsilon^{-1}$ must individually vanish as $\epsilon\rightarrow 0$. In particular,
\begin{align}
0&=\bm{E}_0+\bm{u}_{i0}\times\bm{B}_{0}\\
0&=\bm{E}_0+\bm{u}_{e0}\times\bm{B}_0\\
0&=\bm{u}_{e0}-\bm{u}_{i0}.
\end{align}
The only solution of this system of equations is $\bm{E}_0=-\bm{u}_{i0}\times\bm{B}_0$, $\bm{u}_{e0}=\bm{u}_{i0}$. Thus, $Z_0$ must be contained in the set $S_0$. This verifies the claim.
\\ \\
\noindent\emph{remark:} Physically speaking, the set $S_0$ consists of two-fluid states that are current neutral\cite{Kulsrud_book_1983} ($\bm{u}_e=\bm{u}_i$) and that satisfy the ideal Ohm's Law ($\bm{E}=-\bm{u}_i\times\bm{B}$). Charge neutrality is also enforced because the electron number density $Z_i n_{i\epsilon}+\epsilon^2\epsilon_oq_e^{-1}\nabla\cdot\bm{E}_\epsilon\rightarrow Z_i n_{i0}$ as $\epsilon\rightarrow 0$. Thus, property 2 already provides some evidence that the slow initial conditions in the two-fluid-Maxwell system must be related to MHD. Property 2 also suggests that slow solutions of the two-fluid-Maxwell system have the property that the electric field and the electron fluid velocity are \emph{slaved} to the MHD state variables $(n_i,\bm{u}_i,\bm{B})$, i.e. the former are expressed as functions of the latter. Such slaving relations, which may also be thought of as defining a closure, are commonplace in the theory of slow manifold reduction\cite{Lorenz_1986} and geometric singular perturbation theory\cite{Fenichel_1979}, which forms the theoretical basis underlying the discussion in this section.
\\ \\
Taken together, properties 1 and 2 indicate that a reasonable approach to finding the slow initial conditions is to look for an invariant subset of $Z$-space, $S_\epsilon$, of the form
\begin{align}\label{slow_manifold}
S_\epsilon=&\{(n_i,\bm{u}_i,\bm{B},\bm{u}_e,\bm{E})\mid \bm{u}_e=\bm{u}_{e\epsilon}^*(n_i,\bm{u}_i,\bm{B})\nonumber\\
&\hspace{9em}\bm{E}=\bm{E}_\epsilon^*(n_i,\bm{u}_i,\bm{B})\},
\end{align}
where $\bm{u}_{e\epsilon}^*$ and $\bm{E}_\epsilon^*$ are undetermined  $O(1)$ functions of the MHD state variables $(n_i,\bm{u}_i,\bm{B})$. Keeping in line with the remark below property 2, I will refer to $\bm{u}_{e\epsilon}^*$ and $\bm{E}_\epsilon^*$ as the \emph{slaving functions} of the \emph{slow manifold} $S_\epsilon$. Admittedly, Eq.\,(\ref{slow_manifold}) is nothing more than an ansatz for the set of slow initial conditions. In particular, it is not at all obvious that $S_\epsilon$ needs to exist. Moreover, even if $S_\epsilon$ does exist, that would not imply that $S_\epsilon$ contains \emph{all} of the slow initial conditions -- in fact it is not obvious at this stage that $S_\epsilon$ contains any slow initial conditions whatsoever! Nevertheless, (\ref{slow_manifold}) will prove to be a good ansatz for two reasons. First, it will turn out that the slaving functions have unique asymptotic expansions in powers of $\epsilon$. Second, it will be possible to show formally that the dynamics of two-fluid states in $S_\epsilon$ are indeed slow.

By extending the argument supporting property 2, I will now derive a functional partial differential equation satisfied by the slaving functions in Eq.\,(\ref{slow_manifold}). Suppose that $Z_\epsilon=(n_{i\epsilon},\bm{u}_{i\epsilon},\bm{B}_\epsilon,\bm{u}_{e\epsilon},\bm{E}_\epsilon)$ is a solution of the scaled two-fluid-Maxwell system contained in $S_\epsilon$. Then the electric field and the electron fluid velocity must satisfy the slaving relations
\begin{align}
\bm{u}_{e\epsilon}&=\bm{u}_{e\epsilon}^*(n_{i\epsilon},\bm{u}_{i\epsilon},\bm{B}_\epsilon)\\
\bm{E}_\epsilon&=\bm{E}_\epsilon^*(n_{i\epsilon},\bm{u}_{i\epsilon},\bm{B}_\epsilon),
\end{align}
and $Z_\epsilon$ must be a solution of the scaled two-fluid-Maxwell equations (\ref{ion_momentum})-(\ref{faraday}). An interesting consequence of these two constraints is that the electron momentum equation and the Amp\'ere-Maxwell Law may be written in a manner that does not involve any time derivatives. To see this, first note that the slaving relations imply the partial time derivatives of $\bm{E}_\epsilon$ and $\bm{u}_{e\epsilon}$ are given by
\begin{align}\label{chain_rule_u}
\partial_t\bm{u}_{e\epsilon}&=D_{n_i}\bm{u}_{e\epsilon}^*[\partial_t n_{i\epsilon}]+D_{\bm{u}_i}\bm{u}_{e\epsilon}^*[\partial_t\bm{u}_{i\epsilon}]+D_{\bm{B}}\bm{u}_{e\epsilon}^*[\partial_t\bm{B}_\epsilon]\\
\partial_t\bm{E}_\epsilon&=D_{n_i}\bm{E}_{\epsilon}^*[\partial_t n_{i\epsilon}]+D_{\bm{u}_i}\bm{E}_{\epsilon}^*[\partial_t\bm{u}_{i\epsilon}]+D_{\bm{B}}\bm{E}_{\epsilon}^*[\partial_t\bm{B}_\epsilon].\label{chain_rule_e}
\end{align}
Here the symbol $D$ denotes Fr\'echet derivative (see Appendix \ref{derivatives} if unfamiliar with the Fr\'echet derivative). Next use the ion continuity equation, the ion momentum equation, and Faraday's Law to eliminate the partial time derivatives from the right-hand-sides of Eqs.\,(\ref{chain_rule_u}) and (\ref{chain_rule_e}),
\begin{align}
\partial_t\bm{u}_{e\epsilon}=&-D_{n_i}\bm{u}_{e\epsilon}^*[\nabla\cdot(n_{i\epsilon}\bm{u}_{i\epsilon})]+D_{\bm{u}_i}\bm{u}_{e\epsilon}^*[\dot{\bm{u}}_{i E}(n_{i\epsilon},\bm{u}_{i\epsilon})]\nonumber\\
&-D_{\bm{B}}\bm{u}_{e\epsilon}^*[\nabla\times\bm{E}_\epsilon]-\frac{1}{\epsilon}\frac{q_e}{m_e}\nu Z_iD_{\bm{u}_i}\bm{u}_{e\epsilon}^*[\bm{E}_\epsilon+\bm{u}_{i\epsilon}\times\bm{B}_\epsilon]\\
\partial_t\bm{E}_\epsilon&=-D_{n_i}\bm{E}_{\epsilon}^*[\nabla\cdot(n_{i\epsilon}\bm{u}_{i\epsilon})]+D_{\bm{u}_i}\bm{E}_{\epsilon}^*[\dot{\bm{u}}_{iE}(n_{i\epsilon},\bm{u}_{i\epsilon})]\nonumber\\
&-D_{\bm{B}}\bm{E}_{\epsilon}^*[\nabla\times\bm{E}_\epsilon]-\frac{1}{\epsilon}\frac{q_e}{m_e}\nu Z_iD_{\bm{u}_i}\bm{E}_{\epsilon}^*[\bm{E}_\epsilon+\bm{u}_{i\epsilon}\times\bm{B}_\epsilon]
\end{align}
Here $\nu=m_e/m_i$ is the mass ratio and $\dot{\bm{u}}_{i E}(n_i,\bm{u}_i)=-\bm{u}_{i}\cdot\nabla\bm{u}_i-(m_i n_i)^{-1}\nabla\mathsf{p}_i(n_i)$ is the time derivative of the ion velocity in the absence of electromagnetic forces (``$E$" stands for Euler equation). Because the only temporal derivatives that appear in the electron momentum equation and the Amp\'ere-Maxwell Law are $\partial_t\bm{u}_e$ and $\partial_t\bm{E}$, these manipulations suffice to eliminate all of the time derivatives from these equations.  Moreover, because $Z_\epsilon$ is an arbitrary solution contained in the slow manifold $S_\epsilon$, the time-derivative-free forms of the electron momentum equation and the Amp\'ere-Maxwell Law may be read as the following system of functional partial differential equations for the unknown functionals $\bm{u}_{e\epsilon}^*(n_i,\bm{u}_i,\bm{B})$ and $\bm{E}_\epsilon^*(n_i,\bm{u}_i,\bm{B})$:
\begin{align}
&\frac{1}{\epsilon}\frac{q_e}{m_e}\nu Z_i D_{\bm{u}_i}\bm{u}_{e\epsilon}^*[\bm{E}_\epsilon^*+\bm{u}_i\times\bm{B}]+\frac{1}{\epsilon}\frac{q_e}{m_e}(\bm{E}_\epsilon^*+\bm{u}_{e\epsilon}^*\times\bm{B})\nonumber\\
&=\bm{u}_{e\epsilon}^*\cdot\nabla\bm{u}_{e\epsilon}^*+D_{\bm{u}_i}\bm{u}_{e\epsilon}^*[\dot{\bm{u}}_{iE}]+(m_e n_{e\epsilon}^*)^{-1}\nabla\mathsf{p}_{e}(n_{e\epsilon}^*)\nonumber\\
&\hspace{4em}-D_{n_i}\bm{u}_{e\epsilon}^*[\nabla\cdot(n_i\bm{u}_i)]-D_{\bm{B}}\bm{u}_{e\epsilon}^*[\nabla\times\bm{E}_\epsilon^*]\label{invariance_momentum}\\
&\frac{1}{\epsilon}\mu_o q_e Z_i n_i (\bm{u}_{e\epsilon}^*-\bm{u}_i)=\nabla\times\bm{B}+\mu_o\epsilon_o\frac{q_e}{m_e}\nu Z_i D_{\bm{u}_i}\bm{E}_\epsilon^*[\bm{E}_\epsilon^*+\bm{u}_i\times\bm{B}]\nonumber\\
&-\epsilon\mu_o\epsilon_o\bigg(D_{\bm{u}_i}\bm{E}_\epsilon^*[\dot{\bm{u}}_{iE}]+\bm{u}_{e\epsilon}^*\nabla\cdot\bm{E}_\epsilon^*-D_{n_i}\bm{E}_\epsilon^*[\nabla\cdot(n_i\bm{u}_i)]-D_{\bm{B}}\bm{E}_\epsilon^*[\nabla\times\bm{E}_\epsilon^*]\bigg),\label{invariance_ampere}
\end{align}
where $n_{e\epsilon}^*=Z_i n_i+\epsilon^2\epsilon_oq_e^{-1}\nabla\cdot\bm{E}_\epsilon^*$ is shorthand notation for the electron number density. I will refer to this system of functional PDE as the invariance equations. 

In general the invariance equations, which are nonlinear and involve both functional and ordinary derivatives, are hopelessly difficult to solve. However, if they admit a solution that is smooth in $\epsilon$ and $O(1)$ as $\epsilon\rightarrow 0$, then this solution has the unique asymptotic expansion
\begin{align}
\bm{u}_{e\epsilon}^*&=\bm{u}_{e0}^*+\epsilon \bm{u}_{e1}^*+\epsilon^2\bm{u}_{e2}^*+\dots\label{u_expansion}\\
\bm{E}_\epsilon^*&=\bm{E}_0^*+\epsilon\bm{E}_1^*+\epsilon^2\bm{E}_2^*+\dots,\label{e_expansion}
\end{align}
where the coefficients $\bm{u}_{ek}^*,\bm{E}_k^*$ are most readily obtained by substituting the asymptotic expansions into the invariance equations and then solving order by order. For instance, the leading-order invariance equations ($O(\epsilon^{-1})$ as written) are
\begin{gather}
\frac{q_e}{m_e}\nu Z_i D_{\bm{u}_i}\bm{u}_{e0}^*[\bm{E}_0^*+\bm{u}_i\times\bm{B}]+\frac{q_e}{m_e}(\bm{E}_0^*+\bm{u}_{e0}^*\times\bm{B})=0\\
\mu_o q_e Z_i n_i (\bm{u}_{e0}^*-\bm{u}_i)=0,
\end{gather}
which have the unique solution
\begin{align}
\bm{u}_{e0}^*&=\bm{u}_i\\
\bm{E}_0^*&=-\bm{u}_i\times\bm{B},
\end{align}
representing current neutrality and ideal Ohm's law.
In general, the $n$'th order invariance equation determines uniquely the $n$'th order terms in the asymptotic expansions (\ref{u_expansion}) and (\ref{e_expansion}). In particular, the $O(1)$ invariance equations lead to
\begin{align}
\bm{u}_{e1}^*&=\frac{\mu_o^{-1}\nabla\times\bm{B}}{q_e Z_i n_i}\\
\bm{E}_1^*&=-\frac{\mu_o^{-1}(\nabla\times\bm{B})\times\bm{B}-\nabla(\mathsf{p}_e(Z_i n_i)-\nu Z_i\mathsf{p}_i(n_i))}{q_e Z_i n_i(1+\nu Z_i)},\label{E_one_star}
\end{align}
while the $O(\epsilon)$ invariance equations gives
\begin{align}
\bm{u}_{e2}^*=&-\frac{\rho_{\text{MHD}}\bm{u}_i+\epsilon_o\dot{\bm{E}}_{\text{MHD}}}{q_e Z_i n_i}\\
\bm{E}_2^*=&\frac{(\rho_{\text{MHD}}\bm{u}_i+\epsilon_o \dot{\bm{E}}_{\text{MHD}})\times\bm{B}}{(1+\nu Z_i)q_e Z_i n_i}\nonumber\\
&+\frac{c^2}{\omega_p^2}\bigg((\nabla\times\bm{B})\cdot\nabla\bm{u}_i+(\bm{u}_i\cdot\nabla)\nabla\times\bm{B}\nonumber\\
&\hspace{3em}+(\nabla\times\bm{B}) n_i^{-1}\nabla\cdot(n_i\bm{u}_i)+\nabla\times\nabla\times(\bm{u}_i\times\bm{B})\bigg)
\end{align}
where 
\begin{align}
\rho_{\text{MHD}}=&-\epsilon_o\nabla\cdot(\bm{u}_i\times\bm{B})\label{rho_MHD}\\
\dot{\bm{E}}_{\text{MHD}}=&-\bigg(-\bm{u}_i\cdot\nabla\bm{u}_i+\frac{-\nabla(\mathsf{p}_i(n_i)+\mathsf{p}_e(Z_in_i))+\mu_o^{-1}(\nabla\times\bm{B})\times\bm{B}}{m_in_i(1+\nu Z_i)}\bigg)\times\bm{B}\nonumber\\
&\hspace{8em}-\bm{u}_i\times\nabla\times(\bm{u}_i\times\bm{B})\label{Edot_MHD}
\end{align}
is shorthand notation for the charge density and displacement current given by the ideal MHD model. The higher-order coefficients rapidly become very complicated, but they may be efficiently computed if desired by solving the invariance equations iteratively using a computer algebra system.

Now that the slaving functions have been determined, the time evolution of two-fluid states that are contained in the slow manifold $S_\epsilon$ is easy to determine. Suppose $Z(t)\in S_\epsilon$ is two-fluid trajectory contained in the slow manifold. Combining the ion momentum equation, the ion continuity equation, Faraday's Law, and the slaving relations then implies
\begin{gather}\label{slow_momentum}
m_i n_i (\partial_t\bm{u}_t+\bm{u}_i\cdot\nabla\bm{u}_i)=-\nabla\mathsf{p}_i(n_i)-\frac{1}{\epsilon}q_e Z_i n_i(\bm{E}_\epsilon^*(n_i,\bm{u}_i,\bm{B})+\bm{u}_i\times\bm{B})\\
\partial_t n_i+\nabla\cdot(n_i\bm{u}_i)=0\label{slow_continuity}\\
\partial_t\bm{B}=-\nabla\times\bm{E}_\epsilon^*(n_i,\bm{u}_i,\bm{B}),\label{slow_faraday}
\end{gather}
which clearly gives a closed system of equations that determine the time evolution of the MHD state $(n_i,\bm{u}_i,\bm{B})$. Now, with the time evolution of the MHD state determined, the time evolution of the entire two-fluid state is also determined because two fluid states contained in $S_\epsilon$ have the form $(n_i,\bm{u}_i,\bm{B},\bm{u}_{e\epsilon}^*,\bm{E}_\epsilon^*)$. It is therefore sensible to refer to the system (\ref{slow_momentum})-(\ref{slow_faraday}) as the \emph{slow two-fluid equations}. Observe that the slaving function $\bm{E}_\epsilon^*$ appears in the slow two-fluid equations while $\bm{u}_{e\epsilon}^*$ does not. 

The following facts pertaining to two-fluid configurations that begin in the slow manifold $S_\epsilon$ may now be inferred.
\\ \\
\emph{fact 1:} Two-fluid states that begin on the slow manifold $S_\epsilon$ remain on the slow manifold and evolve on the slow ($O(1)$) timescale. This fact follows from the construction of $S_\epsilon$, which guaranteed $S_\epsilon$ is an invariant set, and from the expressions for the first few terms in the asymptotic expansion for $\bm{E}_\epsilon^*$. Indeed, by substituting the first two terms in the asymptotic expansion of $\bm{E}_\epsilon^*$ into Eq.\,(\ref{slow_momentum}), it is simple to verify that the leading-order truncation of the slow two-fluid equations is given by 
\begin{gather}\label{mhd_recovered_1}
(1+\nu Z_i)m_i n_i (\partial_t\bm{u}_i+\bm{u}_i\cdot\nabla\bm{u}_i)=-\nabla(\mathsf{p}_i(n_i)+\mathsf{p}_e(Z_i n_i))+\mu_o^{-1}(\nabla\times\bm{B})\times\bm{B}\\
\partial_tn_i+\nabla\cdot(n_i\bm{u}_i)=0\\
\partial_t\bm{B}=\nabla\times(\bm{u}_i\times\bm{B}).\label{mhd_recovered_3}
\end{gather}
This shows that the time derivatives $(\partial_t n_i,\partial_t\bm{u}_i,\partial_t\bm{E})$ are each $O(1)$.
\\ \\
\emph{fact 2:} The slow two-fluid equations are equivalent to a formally-exact extended MHD model. That slow dynamics extends ideal MHD is immediately apparent from Eqs.\,(\ref{mhd_recovered_1})-(\ref{mhd_recovered_3}), which of course reproduce the ideal MHD model (with a renormalized ion mass). Interestingly, the first correction to the leading-order slow two-fluid equations is
\begin{gather}
(1+\nu Z_i)m_i n_i (\partial_t\bm{u}_i+\bm{u}_i\cdot\nabla\bm{u}_i)=-\nabla(\mathsf{p}_i(n_i)+\mathsf{p}_e(Z_i n_i))\nonumber\\+[\mu_o^{-1}(\nabla\times\bm{B})-\epsilon\rho_{\text{MHD}}\bm{u}_i-\epsilon\epsilon_o \dot{\bm{E}}_{\text{MHD}}]\times\bm{B}\nonumber\\
\hspace{-11em}-\epsilon (1+\nu Z_i)q_e Z_i n_i\frac{c^2}{\omega_p^2}\bigg((\nabla\times\bm{B})\cdot\nabla\bm{u}_i+(\bm{u}_i\cdot\nabla)\nabla\times\bm{B}\nonumber\\
\hspace{11em}+(\nabla\times\bm{B}) n_i^{-1}\nabla\cdot(n_i\bm{u}_i)+\nabla\times\nabla\times(\bm{u}_i\times\bm{B})\bigg)\label{hall_reproduced_1}\\
\partial_tn_i+\nabla\cdot(n_i\bm{u}_i)=0\\
\partial_t\bm{B}=\nabla\times\left(\bm{u}_i\times\bm{B}+\frac{\epsilon\mu_o^{-1}(\nabla\times\bm{B})\times\bm{B}}{(1+\nu Z_i)q_e Z_i n_i }\right),\label{hall_reproduced_3}
\end{gather}
which is a generalization of Hall MHD\cite{Lighthill_1960,Hameiri_2005}. The reason for the additional terms relative to ordinary Hall MHD is the following. In the ordinary Hall theory, deviations from charge neutrality, the displacement current, and electron inertia are completely ignored. Formally this corresponds to enforcing $\epsilon_o=0$ and $\nu =0$. However, the analysis here makes the weaker assumption $\epsilon_o=O(\epsilon)$ and $\nu=O(1)$, which allows for perturbative deviations from exact charge neutrality, as well as perturbative contributions to the transverse electric field and the full effects of finite electron inertia. This explains the extra terms in Eqs.\,(\ref{hall_reproduced_1})-(\ref{hall_reproduced_3}). If a subsidiary ordering were introduced, or if I had instead assumed $\epsilon_o=O(\epsilon^k)$, $\nu=O(\epsilon)$ with $k>1$, Eqs.\,(\ref{hall_reproduced_1})-(\ref{hall_reproduced_3}) would be identical to Hall MHD. In particular, all terms proportional to $\epsilon$ in the momentum equation would vanish. However, seeing as the ordering used in this analysis is less strict than the conventional ordering, it is entirely possible that the generalized Hall MHD equations given in Eqs.\,(\ref{hall_reproduced_1})-(\ref{hall_reproduced_3}) are more accurate than conventional Hall MHD, especially in situations where deviations from charge neutrality are moderately important. 

\section{The Hamiltonian structure governing slow two-fluid dynamics\label{Hamiltonian_theory}}
The perturbative solution of the invariance equations (\ref{invariance_momentum})-(\ref{invariance_ampere}) presented in the previous section gives suggestive evidence that one possible mechanism for the emergence of MHD behavior within the two-fluid model is the existence of a slow manifold. However, one drawback of the perturbative solution is that high-order contributions to the slaving functions $\bm{E}_\epsilon^*,\bm{u}_{e\epsilon}^*$, which define the slow manifold, are extremely difficult to calculate. This makes it difficult to distinguish between properties that the slow two-fluid equations genuinely possess, and properties that only particular truncations of the slow two-fluid equations possess. For instance, while the leading-order truncation of the slow two-fluid equations (i.e. MHD) gives a system of PDE that is first-order in both space and time, the Hall MHD truncation at next-to-leading order involves second-order derivatives in space. It is therefore entirely unclear what category of PDE the slow two-fluid equations fall into. In fact it is possible, if not likely, that the appearance of high-order space derivatives in high-order truncations of the slow two-fluid equations is merely an artifact of unwittingly expanding nonlocal operators in powers of $\epsilon$.

There is one set of properties possessed by the full slow two-fluid system that \emph{can} be understood in a rather complete sense. These special properties pertain to the Hamiltonian structure of the slow dynamics. This section will show that the Poisson bracket governing slow two-fluid dynamics may be obtained in closed form. Using this expression, it will be possible to deduce further closed form expressions for some of the conservation laws possessed by the all-orders slow two-fluid system. In addition, this result will lead to a convenient and practically useful method for obtaining \emph{conservative} truncations of the slow two-fluid system. Because the data consisting of the Poisson bracket and the Hamiltonian functional completely determines the slow two-fluid equations, the problem of truncating while preserving the conservative properties of the slow dynamics is reduced to truncating the Hamiltonian. 

In order to uncover the slow two-fluid system's Hamiltonian structure, it is easiest to start from the phase space variational principle governing the full two-fluid-Maxwell system, which is embodied by the phase space Lagrangian (\ref{phase_space_lagrangian}). Given a solution of the two-fluid Maxwell system, the phase space variational principle states that an arbitrary variation of the action $\mathcal{S}=\int L\,dt$ around that solution vanishes, $\delta \mathcal{S}=0$. In particular, if the solution is contained within the slow manifold $S_\epsilon$, any variation of the action \emph{that does not leave the slow manifold} vanishes. In other words, the action $\mathcal{S}^*$ obtained by restricting $\mathcal{S}$ to curves that are contained in $S_\epsilon$ has as critical points solutions of the slow two-fluid equations. This implies that the slow two-fluid system \emph{inherits} a phase space variational principle from the full two-fluid Maxwell system. This slow two-fluid variational principle represents the first crucial step toward obtaining a closed for expression for the slow two-fluid Poisson bracket.

Explicitly, the slow two-fluid action is given by $\mathcal{S}^*=\int L^*\,dt$, where the slow two-fluid Lagrangian is, in accordance with the previous paragraph, given by
\begin{align}
L^*=&\int (Z_i n_i +\epsilon^2 \epsilon_o q_e^{-1}\nabla\cdot\bm{E}_\epsilon^*)(m_e \bm{u}_{e\epsilon}^*+\epsilon^{-1}q_e \bm{A})\cdot\bm{v}_e\,d^3\bm{x}+\int n_i (m_i\bm{u}_i-\epsilon^{-1}Z_i q_e \bm{A})\cdot\bm{v}_i\,d^3\bm{x}\nonumber\\
&\hspace{10em}-\int\epsilon \epsilon_o\bm{E}_\epsilon^*\cdot\dot{\bm{A}}\,d^3\bm{x}-\mathcal{H}^*,
\end{align}
and the slow two-fluid Hamiltonian is given by
\begin{align}
\mathcal{H}^*&=\frac{1}{2}\int m_i n_i |\bm{u}_i|^2\,d^3\bm{x}+\frac{1}{2}\int m_e (Z_i n_i +\epsilon^2 \epsilon_o q_e^{-1}\nabla\cdot\bm{E}_\epsilon^*)|\bm{u}_{e\epsilon}^*|^2\,d^3\bm{x}\nonumber\\
&+\int n_i \,\mathcal{U}_i(n_i)\,d^3\bm{x}+\int (Z_i n_i +\epsilon^2 \epsilon_o q_e^{-1}\nabla\cdot\bm{E}_\epsilon^*)\,\mathcal{U}_e(Z_i n_i +\epsilon^2 \epsilon_o q_e^{-1}\nabla\cdot\bm{E}_\epsilon^*)\,d^3\bm{x}\nonumber\\
&\hspace{5em}+\frac{1}{2}\int\epsilon \epsilon_o|\bm{E}_\epsilon^*|^2\,d^3\bm{x}+\frac{1}{2}\int \mu_o^{-1}|\bm{B}|^2\,d^3\bm{x}.
\end{align}
Notice that in these expressions the velocity $\bm{u}_e$ has been replaced with the slaving function $\bm{u}_{e\epsilon}^*$ while the velocity $\bm{v}_e$ has not. The reason for this is that the constraint imposed by restricting to the slow manifold $S_\epsilon$ involves only \emph{Eulerian} quantities. The Lagrangian configuration maps for both electrons and ions are completely unconstrained on the slow manifold. Of course, one may worry that if the velocity variable $\bm{u}_e$ is constrained, then there might have to be a corresponding non-holonomic constraint on the electron configuration map. This is faulty reasoning because here we are working with a phase space Lagrangian. The paths that we vary in the action $\mathcal{S}$ do not generally satisfy $\bm{u}_e=\bm{v}_e$, even though this relationship must hold along any solution of the Euler-Lagrange equations.

Using the phase space Lagrangian $L^*$ it is possible in principle to identify the Hamiltonian structure underlying the slow two-fluid dynamics. In fact, the Poisson bracket associated with $L^*$ is encoded in the part of $L^*$ that is linear in the velocities $(\bm{v}_i,\bm{v}_e,\dot{\bm{A}},\dot{\bm{u}}_i)$ -- the so-called ``symplectic part'' of $L^*$ (see Section II.C in Ref.\,\onlinecite{Cary_2009}). Littlejohn\cite{Littlejohn_1981,Littlejohn_1983} gives a lucid discussion of this point, albeit in a finite-dimensional context, in his seminal work on the Hamiltonian formulation of guiding center motion. However, the situtation is more complicated than it seems to be at first glance.  Because the symplectic part of $L^*$ involves the slaving functions $\bm{E}_\epsilon^*,\bm{u}_{e\epsilon}^*$, the Poisson bracket associated with $L^*$ is very complicated. In fact, as a result of the slaving functions being known only as formal infinite series in $\epsilon$, the Poisson bracket must also be a formal infinite series in $\epsilon$. Thus, while the Hamiltonian structure for the slow dynamics is just in reach, the sought-after closed-form expression for the Poisson bracket still lies waiting on the other side of a theoretical chasm that has yet to be crossed.

A closed-form expression for the slow dynamics' Poisson bracket may be obtained by employing some ideas from symplectic geometry. The specific ideas that are necessary are described well by Littlejohn in Ref.\,\onlinecite{Littlejohn_1982}. From Littlejohn's work, a procedure may be extracted that greatly simplifies the slow system's Poisson bracket. The specific details of this procedure, being perhaps more abstract than the present discussion requires, are contained in Appendix \ref{transformation_theory}. However, the essential idea behind the procedure is simple to describe and to understand. The goal is merely to find a near-identity transformation of the slow manifold $S_\epsilon$ such that the linear terms in the transformed Lagrangian truncate at a finite order in $\epsilon$. If such a transformation can be found, it would lead to an easily-computable closed-form expression for the Poisson bracket. The only price of going in this direction, aside from the labor involved, would be a somewhat more complicated Hamiltonian. As explained in the Appendix, existence of a transformation with the desired properties is ensured by a straightforward application of a geometric tool known as Moser's trick (see Ch. 3 in Ref.\,\onlinecite{Abraham_2008}).

One transformation that leads to an $O(1)$ truncation of the symplectic part of $L^*$ is given by
\begin{align}
n_i&=\left(1+\epsilon ^2\frac{\nu Z_i}{1+\nu Z_i}\frac{\nabla\cdot(\bm{\mathsf{u}}\times\bm{\mathsf{B}})}{q_e Z_i \mathsf{n}}\right) \mathsf{n}+O(\epsilon^3)\\
\bm{u}_i&=\bm{\mathsf{u}}+\epsilon (\bm{\varepsilon}^{-1}-1-\bm{\mathsf{b}}\bm{\mathsf{b}})\cdot\bigg(\frac{\mathsf{v}_A^2}{c^2}\frac{\tilde{\bm{\mathsf{E}}}\times\bm{\mathsf{B}}}{|\bm{\mathsf{B}}|^2}+\frac{\nu Z_i}{1+\nu Z_i}\frac{\tilde{\bm{\mathsf{J}}}}{q_e Z_i \mathsf{n}}\bigg)+O(\epsilon^3)\\
\bm{A}&=\bm{\mathsf{A}}-\epsilon^2\Bigg(\frac{m_i}{q_e Z_i}\left((1-\nu Z_i)\frac{\mathsf{v}_A^2}{c^2}\frac{\tilde{\bm{\mathsf{E}}}\times\bm{\mathsf{B}}}{|\bm{\mathsf{B}}|^2}+\frac{\nu Z_i}{1+\nu Z_i}\frac{\tilde{\bm{\mathsf{J}}}}{q_e Z_i \mathsf{n}}\right)\nonumber\\
&\hspace{12.5em}+\frac{1}{1+\nu Z_i}\nabla\left(\frac{\bm{\mathsf{A}}\cdot\epsilon_o \bm{\mathsf{u}}\times\bm{\mathsf{B}} }{q_e Z_i \mathsf{n}}\right)\Bigg)+O(\epsilon^3),
\end{align} 
where
\begin{align}
\bm{\varepsilon}&=1+\epsilon \frac{c^2}{\mathsf{v}_A^2}(1-\bm{\mathsf{b}}\bm{\mathsf{b}})\\
\tilde{\bm{\mathsf{E}}}&=-\bm{\mathsf{u}}\times\bm{\mathsf{B}}+\epsilon\bm{\mathsf{E}}_1^*\\
\tilde{\bm{\mathsf{J}}}&=\mu_o^{-1}\nabla\times\bm{\mathsf{B}}-\epsilon\uprho_{\text{MHD}}\bm{\mathsf{u}}-\epsilon \epsilon_o \dot{\bm{\mathsf{E}}}_{\text{MHD}},
\end{align}
and $\mathsf{v}_A^2,\bm{\mathsf{E}}_1^*,\dot{\bm{\mathsf{E}}}_{\text{MHD}},\uprho_{\text{MHD}}$ are merely $v_A^2,\bm{E}_1^*,\dot{\bm{E}}_{\text{MHD}},\rho_{\text{MHD}}$ evaluated using the transformed variables $\mathsf{n},\bm{\mathsf{u}},\bm{\mathsf{B}}=\nabla\times\bm{\mathsf{A}}$; recall Eqs.(\ref{scalars}), (\ref{E_one_star}), (\ref{rho_MHD}), and (\ref{Edot_MHD}).
The transformed Lagrangian is given by
\begin{align}\label{transformed_L}
\bar{L}^*&=\frac{\nu Z_i}{1+\nu Z_i}\int \mathsf{m}\mathsf{n} \bm{\mathsf{u}}\cdot\bm{\mathsf{v}}_e\,d^3\bm{x}+\frac{1}{\epsilon}\int q_e Z_i\mathsf{n}\bm{\mathsf{A}}\cdot\bm{\mathsf{v}}_e\,d^3\bm{x} \nonumber\\
&+\frac{1}{1+\nu Z_i}\int \mathsf{m}\mathsf{n}\bm{\mathsf{u}}\cdot\bm{\mathsf{v}}_i\,d^3\bm{x}-\frac{1}{\epsilon}\int q_e Z_i\mathsf{n}\bm{\mathsf{A}}\cdot\bm{\mathsf{v}}_i\,d^3\bm{x}-\bar{\mathcal{H}}^*
\end{align}
where the transformed Hamiltonian is given by $\bar{\mathcal{H}}^*=\bar{\mathcal{H}}^*_0+\epsilon \bar{\mathcal{H}}^*_1+\epsilon^2\bar{\mathcal{H}}^*_2+O(\epsilon^3)$, with
\begin{align}
\bar{\mathcal{H}}^*_0&=\frac{1}{2}\int \mathsf{m} \mathsf{n}|\bm{\mathsf{u}}|^2\,d^3\bm{x}+\int \mathsf{n}\,\mathcal{U}_i(\mathsf{n})\,d^3\bm{x}+\int Z_i\mathsf{n}\,\mathcal{U}_e(Z_i\mathsf{n})\,d^3\bm{x}+\frac{1}{2}\int \mu_o^{-1} |\bm{\mathsf{B}}|^2\,d^3\bm{x}\\
\bar{\mathcal{H}}^*_1&=-\frac{1}{2}\int \mathsf{m}\mathsf{n}\frac{v_A^2}{c^2}|\bm{\mathsf{u}}_\perp|^2\\
\bar{\mathcal{H}}^*_2&=\frac{1}{2}\int \mathsf{m}\mathsf{n}\frac{v_A^4}{c^4}|\bm{\mathsf{u}}_\perp|^2\,d^3\bm{x}-(1-\nu Z_i)\int\mathsf{m}\mathsf{n}\frac{v_A^2}{c^2}\bm{\mathsf{u}}_\perp\cdot\frac{ \mu_o^{-1}\nabla\times\bm{\mathsf{B}}}{(1+\nu Z_i)q_e Z_i\mathsf{n}}\,d^3\bm{x}\nonumber\\
&-\frac{1}{2}\int \frac{c^2}{\omega_p^2}\mu_o^{-1}|\nabla\times\bm{\mathsf{B}}|^2\,d^3\bm{x}+\int\mathsf{m}\mathsf{n}\frac{v_A^2}{c^2}\bm{\mathsf{u}}_\perp\cdot\frac{\bm{\mathsf{u}}_{De}+\nu Z_i\bm{\mathsf{u}}_{Di}}{1+\nu Z_i}\,d^3\bm{x}.
\end{align}
Here the convenient shorthand notations $\bm{\mathsf{u}}_\perp=(1-\bm{\mathsf{b}}\bm{\mathsf{b}})\cdot\bm{\mathsf{u}}$ and $\mathsf{m}=(1+\nu Z_i)m_i$ have been introduced, as well as the so-called diamagnetic drift velocities
\begin{align}
\bm{\mathsf{u}}_{Di}&=\frac{\nabla\mathsf{p}_i(\mathsf{n})\times \bm{\mathsf{B}}}{q_e Z_i \mathsf{n}|\bm{\mathsf{B}}|^2 }\\
\bm{\mathsf{u}}_{De}&=\frac{-\nabla \mathsf{p}_e(Z_i\mathsf{n})\times\bm{\mathsf{B}}}{q_e Z_i\mathsf{n}|\bm{\mathsf{B}}|^2}.
\end{align}
It is important to emphasize here that even though the displayed expressions for the transformation $(n_i,\bm{u}_i,\bm{A})\mapsto (\mathsf{n},\bm{\mathsf{u}},\bm{\mathsf{A}})$ contain only finitely-many terms, the full transformation contains infinitely-many terms. As explained in Appendix \ref{transformation_theory}, all of these terms may be calculated in a systematic manner, and they ensure that the symplectic part of the transformed phase space Lagrangian is displayed entirely in Eq.\,(\ref{transformed_L}); higher-order corrections to the symplectic part are zero in the ``nice'' coordinate system on $S_\epsilon$. It is also worth mentioning that the transformation calculated here is not the only one that leads to a simplified symplectic part in the phase space Lagrangian. 

The Poisson bracket associated with $\bar{L}^*$ may be found using standard techniques. In particular, the computation may be done by inverting the Lagrange tensor\cite{Goldstein_2002} associated with the symplectic part of $\bar{L}^*$ as is done for various kinetic systems in Ref.\,\onlinecite{Burby_thesis_2015}. Proceeding in this manner is useful because it ensures that the Poisson bracket satisfies the Jacobi identity; direct verification of the Jacobi identity as in Ref.\,\onlinecite{Abdelhamid_2015} is not necessary. I will merely report the result of this calculation here. Given two arbitrary functionals $\mathcal{G}(\mathsf{n},\bm{\mathsf{u}},\bm{\mathsf{B}})$ and $\mathcal{H}(\mathsf{n},\bm{\mathsf{u}},\bm{\mathsf{B}})$, their Poisson bracket is given by
\begin{align}\label{bracket}
\{\mathcal{G},\mathcal{H}\}=&\int \frac{1}{\mathsf{m}}\left\{\left(\nabla\frac{\delta \mathcal{G}}{\delta \mathsf{n}}\right)\cdot \frac{\delta\mathcal{H}}{\delta\bm{\mathsf{u}}}-\left(\nabla\frac{\delta \mathcal{H}}{\delta \mathsf{n}}\right)\cdot \frac{\delta\mathcal{G}}{\delta\bm{\mathsf{u}}}\right\}\,d^3\bm{x}\nonumber\\
+&\int\frac{1}{\mathsf{m}\mathsf{n}}\bm{\mathsf{B}}\cdot\left\{\frac{\delta\mathcal{G}}{\delta\bm{\mathsf{u}}}\times\nabla\times\frac{\delta\mathcal{H}}{\delta\bm{\mathsf{B}}}-\frac{\delta\mathcal{H}}{\delta\bm{\mathsf{u}}}\times\nabla\times\frac{\delta\mathcal{G}}{\delta\bm{\mathsf{B}}}\right\}\,d^3\bm{x}\\
+&\int \frac{1}{\mathsf{m}\mathsf{n}}\nabla\times\bm{\mathsf{u}}\cdot\left\{\frac{\delta\mathcal{G}}{\delta\bm{\mathsf{u}}}\times\frac{\delta\mathcal{H}}{\delta\bm{\mathsf{u}}}\right\}\,d^3\bm{x}\nonumber\\
-\epsilon&\int\frac{1}{\mathsf{m}\mathsf{n}}\bm{\mathsf{B}}\cdot\left\{\left(\frac{|\bm{\mathsf{B}}|}{\omega_{ce}}+\frac{|\bm{\mathsf{B}}|}{\omega_{ci}}\right)\left(\nabla\times\frac{\delta\mathcal{G}}{\delta\bm{\mathsf{B}}}\right)\times\left(\nabla\times\frac{\delta\mathcal{H}}{\delta\bm{\mathsf{B}}}\right)\right\}\,d^3\bm{x}\nonumber\\
+\epsilon^2&\int \frac{1}{\mathsf{m}\mathsf{n}}\nabla\times\bm{\mathsf{u}}\cdot\left\{\nu Z_i \frac{|\bm{\mathsf{B}}|^2}{\omega_{ci}^2}\left(\nabla\times\frac{\delta\mathcal{G}}{\delta\bm{\mathsf{B}}}\right)\times\left(\nabla\times\frac{\delta\mathcal{H}}{\delta\bm{\mathsf{B}}}\right)\right\}\,d^3\bm{x}
\end{align} 
Because this expression does not involve any infinite series, it represents the main result that was meant to be obtained in this section.

With a closed-form expression for the slow system's Poisson bracket in hand, it is now possible to deduce some general properties of the slow two-fluid system that are independent of truncation order. First and foremost, it is now clear that the slow dynamics possess \emph{Hamiltonian} truncations with any desired level of accuracy. Such Hamiltonian truncations are obtained by using the full Poisson bracket in Eq.\,(\ref{bracket}) while retaining only finitely-many terms in the expansion of the transformed Hamiltonian $\bar{\mathcal{H}}^*=\bar{\mathcal{H}}^*_0+\epsilon \bar{\mathcal{H}}^*_1+\epsilon^2 \bar{\mathcal{H}}^*_2+\dots$. These truncations of the slow two-fluid system are superior to naive truncations performed at the level of the equations of motion because they ensure that artificial dissipation is not introduced by truncation. An obvious manifestation of this fact is that any Hamiltonian truncation of the slow dynamics will automatically conserve the truncated Hamiltonian exactly. This follows from the antisymmetry of the Poisson bracket. All Hamiltonian truncations of the slow two-fluid system also possess less obvious conservation laws. Notably, they all possess a pair of circulation invariants, which I will now describe. 

For the sake of describing the circulation invariants precisely, let $\mathfrak{H}=\bar{\mathcal{H}}^*_0+\epsilon \bar{\mathcal{H}}^*_1+\epsilon^2 \bar{\mathcal{H}}^*_2+\dots+\bar{\mathcal{H}}^*_n$ denote the $n$'th order truncation of the transformed Hamiltonian, where $n$ is an arbitrary non-negative integer. The first circulation invariant, which is associated with motion of the ions, is given by
\begin{align}
C_i=\oint_{\ell_i} \bm{\mathsf{A}}\cdot d\bm{x}-\epsilon\frac{ m_i}{q_e Z_i}\oint_{\ell_i} \bm{\mathsf{u}}\cdot d\bm{x}.
\end{align}
Here $\ell_i$ is a closed loop that moves with the velocity
\begin{align}
\bm{\mathsf{v}}_i^*=\frac{1}{\mathsf{m}\mathsf{n}}\frac{\delta\mathfrak{H}}{\delta\bm{\mathsf{u}}}-\epsilon\frac{\nu Z_i}{1+\nu Z_i}\frac{1}{q_e Z_i \mathsf{n}}\nabla\times\frac{\delta\mathfrak{H}}{\delta\bm{\mathsf{B}}}.
\end{align}
As the notation suggests, the velocity $\bm{\mathsf{v}}_i^*$ is approximately equal to the ion fluid velocity. It is well-known that the barotropic two-fluid Maxwell system has an analogous circulation invariant, which is equivalent to the circulation of the ion canonical momentum. Actually, the invariant $C_i$ is the same quantity, merely restricted to the slow manifold. The second circulation invariant, which is associated with motion of electrons, is given by
\begin{align}
C_e=\int_{\ell_e}\bm{\mathsf{A}}\cdot d\bm{x}+\epsilon \nu Z_i\frac{m_i}{q_e Z_i}\int_{\ell_e}\bm{\mathsf{u}}_e\cdot d\bm{x}.
\end{align}
Here $\ell_e$ is an \emph{arbitrary} (not necessarily closed) curve that moves with the velocity
\begin{align}
\bm{\mathsf{v}}_e^*=\frac{1}{\mathsf{m}\mathsf{n}}\frac{\delta\mathfrak{H}}{\delta\bm{\mathsf{u}}}+\epsilon\frac{1}{1+\nu Z_i}\frac{1}{q_e Z_i \mathsf{n}}\nabla\times\frac{\delta\mathfrak{H}}{\delta\bm{\mathsf{B}}}.
\end{align} 
Just like $C_i$, $C_e$ may be interpreted as an invariant inherited from the two-fluid-Maxwell system.

\section{Discussion\label{discussion}}
This article has put forth a new conceptual framework for understanding MHD and its relationship with ideal two-fluid theory. The key insight that underlies this new perspective is that the two-fluid theory, when appropriately scaled, admits a formally-exact single-fluid closure. At leading order, the closure reproduces ideal MHD. At higher orders, the closure leads to new or modified exteneded MHD models. Notably, the latter allow for arbitrary mass ratio as well as perturbative deviations from exact charge neutrality. (Previous extended MHD models enforce strict charge neutrality; the closure here only enforces charge neutrality at the leading (MHD) order).

The full (i.e. all-orders) single-fluid closure is given as the solution of a certain functional PDE, Eqs.\,(\ref{invariance_momentum})-(\ref{invariance_ampere}), which can be solved in an asymptotic sense without essential difficulty. Moreover, the solution of the functional PDE has an appealing geometric interpretation as an invariant submanifold in the infinite-dimensional two-fluid phase space, known as a slow manifold. Two-fluid motions that begin within the slow manifold remain within the slow manifold as time passes. These special solutions of the two-fluid system evolve as if they were governed by a small perturbation of the ideal MHD system. They are free of high-frequency oscillations that otherwise generally occur at the cyclotron and plasma frequency time scales. One mechanism by which a two-fluid plasma may exhibit emergent MHD behavior is therefore initialization of the plasma so that its mechanical state lies within the slow manifold.

A deep corollary of the geometric interpretation of the single-fluid closure is that the closure actually inherits a Hamiltonian structure from the two-fluid model. Using infinite-dimensional Lie transforms, this Hamiltonian structure, which is encoded in the form of a Poisson bracket, may be found in closed form. Therefore, strictly dissipation-free truncations of the single-fluid closure may be obtained by simply truncating an asymptotic expansion for the single-fluid Hamiltonian. Because the single-fluid Hamiltonian is the restriction of the two-fluid Hamiltonian to the slow manifold, an asymptotic expansion of the single-fluid Hamiltonian is readily computable. All such Hamiltonian truncations possess a pair of interesting independent circulation invariants, in addition to the expected invariants associated with space- and time-translation symmetry. This last fact shows that the observations in Ref.\,\onlinecite{Lingam_2015} pertaining to pairs of circulation invariants in some extended MHD models apply much more generally.

It is insightful to compare and contrast the results in this article with previous work on ideal and extended MHD. First consider Ref.\,\onlinecite{Charidakos_2014}, where a two-fluid action was used in conjuction with asymptotic methods to deduce action principles for a selection of previously-known extened MHD models.  The heart and soul of the method used in the present work to obtain the single-fluid Hamiltonian structure may be traced back to this Reference. To be more precise, the present work demonstrates that the basic principle underlying Ref.\,\onlinecite{Charidakos_2014}, namely that of applying asymptotic methods directly to a two-fluid variational principle, may be extended to identify the Hamiltonian structure of the all-orders single-fluid closure of the two-fluid system. The extension involves, amongst some other technical details, employing a two-fluid phase-space variational principle instead of a configuration space variational principle, as well as applying infinite-dimensional Lie transforms to find a closed-form expression for the all-orders Poisson bracket.

There are several notable differences between the present work and that of Ref.\,\onlinecite{Charidakos_2014}. Most importantly, while the present work studies an all-orders single-fluid closure, Ref.\,\onlinecite{Charidakos_2014} is not concerned with establishing all-orders results. As was explained earlier in Section \ref{Hamiltonian_theory}, studying properties of the all-orders theory is important in order to distinguish between phenomena genuinely present in the full single-fluid closure (such as circulation invariants) and phenomena that appear only in particular low-order truncations of the closure (such as the impossibility of magnetic reconnection). On the other hand, given that the primary goal of Ref.\,\onlinecite{Charidakos_2014} was to establish new results on previously-established extended MHD models, there was no particular need for an all-orders theory in Ref.\,\onlinecite{Charidakos_2014}.

Next consider Ref.\,\onlinecite{Abdelhamid_2015}, which identifies a Poisson bracket structure for the strictly-neutral two-fluid system derived by L\"ust.\cite{Lust_1959} It is conceptually satisfying to observe that the all-orders Poisson bracket given here in Eq.\,(\ref{bracket}) is equivalent to the bracket given in Eq.\,(29) of Ref.\,\onlinecite{Abdelhamid_2015} after making the identifications
\begin{gather*}
\bm{\mathsf{u}}=\bm{V}\quad\bm{\mathsf{B}}=\bm{B}^*\quad \mathsf{m}\mathsf{n}=\rho\quad \mu_o m_i\mathsf{n}\frac{c^2}{\omega_{pe}^2}=d_e^2\\
\sqrt{\mu_o m_i \mathsf{n}}\left(\frac{c}{\omega_{pi}}-\sqrt{\nu Z_i}\frac{c}{\omega_{pe}}\right)=d_i.\\
\end{gather*} 
The reason that the $d_i$ in Ref.\,\onlinecite{Abdelhamid_2015} must be identified with a quantity that is not merely proportional to $c/\omega_{pi}$ is that Ref.\,\onlinecite{Abdelhamid_2015} assumes that the mass ratio is small, while in this work the mass ratio $\nu=O(1)$. Apparently the bracket underlying L\"ust's model has a significance that extends beyond the strictly-neutral two-fluid system. Moreover, invoking the results of Ref.\,\onlinecite{Lingam_2015}, it must therefore be true that there is a simple relationship between the all-orders bracket and the bracket\cite{Yoshida_2013} for Hall MHD (see Ref.\,\onlinecite{Hazeltine_1987} for the first bracket accounting for Hall physics, albeit in a reduced setting). Indeed, Ref.\,\onlinecite{Lingam_2015} shows that there is a simple transformation that maps L\'ust's bracket into the Hall MHD bracket. It turns out that the appropriate transformation for the all-orders model is $\bm{\mathsf{A}}\mapsto\bar{\bm{\mathsf{A}}} $, where
\begin{align*}
\bar{\bm{\mathsf{A}}}=\bm{\mathsf{A}}+\epsilon\frac{\nu Z_i}{q_e Z_i}m_i\bm{\mathsf{u}}.
\end{align*}

Looking ahead, it is interesting to consider questions that one would not have asked without this article's newfound phase-space-geometric interpretation of MHD. Perhaps the first question that naturally raises is that of convergence of the perturbative solution of the functional PDE (\ref{invariance_momentum})-(\ref{invariance_ampere}). By summing all of the terms in the asymptotic solution of this equation, does the result converge and represent a truly- (not just formally-) exact single-fluid closure? If the series does not converge, does that mean that an exact single-fluid closure does not exist? I conjecture that the likely answers to these questions are ``no'' and ``yes, but there may as well be one.'' This conjecture is based on what happens in finite dimensions. In the finite-dimensional setting\cite{Kristiansen_2016,MacKay_2004,Neishtadt_1987} (actually Ref.\,\onlinecite{Kristiansen_2016} handles systems with finitely-many slow variables and infinitely-many fast variables!), it has been shown that when the PDE defining a slow manifold in a Hamiltonian system can be solved perturbatively, there actually exists a so-called \emph{almost-invariant set} that is approximated well by truncations of the slow manifold. When solutions start within the almost invariant set, they remain within the almost invariant set, and therefore close to the truncated slow manifold, for exponentially-long periods of time. Extension of these finite-dimensional results to infinite dimensions is certainly a non-trivial task, but one that would have deep implications for the behavior of solutions to a variety of physical models, in plasma physics and elsewhere. Actually, some work in this direction has already been carried out. In Ref.\,\onlinecite{Vanneste_2004}, Vanneste uses exponential asymptotics to show that while the $3D$ Boussinesq equations do not admit an exact quasigeostrophic closure, particular solutions that begin on an approximate slow manifold only deviate from the slow manifold by an exponentially-small amount. 

A second question to ask is ``what about the two-fluid states that do not lie within the slow manifold?" This is an important question because plasmas that live on the slow manifold are extremely lucky; most two-fluid states are not contained within the slow manifold. While this is a difficult question, intuition suggests that the most general motion of a two-fluid plasma bears very little resemblance to the predictions given by MHD. However, there is reason to be hopeful that two-fluid plasmas that begin near, but not exactly on the slow manifold still feel the influence of the MHD model. It is not difficult to show that two-fluid dynamics near the slow manifold generally decompose into a slow drift along the slow manifold and a rapid oscillation transverse to the slow manifold. In the regime $\rho/L\ll 1$, the transverse oscillations describe the dynamics of the Langmuir oscillation and light waves (with the well-known modification to the free-space dispersion relation). By generalizing the oscillation center theory of Refs.\,\onlinecite{Cary_1981,Cary_pond_1981} to infinite dimensions, it may be possible to compute the ponderomotive effect of the transverse oscillations on the drift motion along the slow manifold. If this can be done within the context of the Hamiltonian formalism, there will be an adiabatically-invariant field related to the wave action of light waves and Langmuir waves. This adiabatic invariant will couple the averaged transverse dynamics to the MHD-like single-fluid closure \emph{via} some kind of effective potential. This topic will be thoroughly explored in future work. 

\section{Acknowledgements}
This research was supported by the U.S. Department of Energy Fusion Energy Sciences Postdoctoral Research Program administered by the Oak Ridge Institute for Science and Education (ORISE) for the DOE. ORISE is managed by Oak Ridge Associated Universities (ORAU) under DOE contract number DE-AC05-06OR23100. All opinions expressed in this paper are the author's and do not necessarily reflect the policies and views of DOE, ORAU, or ORISE.
%%%%%%%%%%%%%%%%%%%%%%%%%%%%%%%%%%%%%%%%%%%%%%%%%%%%%%%%%%%%%%%%%%%

\appendix
\section{Fr\'echet derivatives\label{derivatives}}
Given a vector space $W$ of (possibly vector-valued) fields $\psi\in W$, it is natural to wonder how (possibly nonlinear) functionals $F: W\rightarrow V$ that take values in a vector space $V$ may be differentiated. One natural answer is provided by the Fr\'echet derivative, $DF$. (For more discussion see Ref.\,\onlinecite{Abraham_2008}.) In order to define $DF$, it is necessary to introduce the technical assumption that $W$ and $V$ are complete normed spaces, i.e. Banach spaces. The norms on $W$ and $V$ will be denoted $|\cdot|_W$ and $|\cdot|_V$.  Like $F$, $DF$ is a (possibly nonlinear) functional of $\psi$. However, the value of $DF$ at $\psi$, $DF(\psi)$, is not an element of $V$.  Instead $DF(\psi)$ is a linear operator that maps $W$ into $V$. We may write $DF(\psi):W\rightarrow V$. This particular operator is defined as the linearization of the map $F$ about the point $\psi$, i.e. if $\delta\psi\in W$ represents an infinitesimal displacement of the field $\psi$, then $DF(\psi)$ is the unique linear operator such that
\begin{align}
\lim_{\delta\psi\rightarrow 0}\frac{|F(\psi+\delta\psi)-F(\psi)-DF(\psi)[\delta\psi]|_V}{|\delta \psi|_W}=0.
\end{align}
Of course, there is no guarantee that $DF(\psi)$ exists. If it does, it is unique and it is said that $F$ is Fr\'echet differentiable at $\psi$.

 The most basic property satisfied by the Fr\'echet derivative is the chain rule. If $F:W\rightarrow V$ and $G:V\rightarrow U$, then
\begin{align}
D(G\circ F)(\psi)[\delta\psi]=DG(F(\psi))[DF(\psi)[\delta\psi]],
\end{align}
whenever the derivatives on the right-hand-side exist. When context suggests where the Fr\'echet derivative is to be evaluated, it is convenient to suppress the nonlinear argument of $DF$, i.e. $DF[\delta\psi]$ may sometimes be written instead of $DF(\psi)[\delta\psi]$. Using this convention, the chain rule may be written as
\begin{align}
D(G\circ F)[\delta\psi]=DG[DF[\delta\psi]],
\end{align}
and it is now necessary to be mindful that $DG$ is evaluated at $F(\psi)$ and $DF$ is evaluated at $\psi$.
A useful consequence of the Chain rule is that the Fr\'echet derivative may be computed using the formula
\begin{align}
DF(\psi)[\delta\psi]=\frac{d}{d\epsilon}\bigg|_0 F(\psi+\epsilon\delta\psi).
\end{align}

As an example, let $\bm{u}$ and $\bm{B}$ be square-integrable vector fields on $\mathbb{R}^3$ and set $F(\bm{u},\bm{B})=\bm{u}\times\bm{B}$. It is not difficult to show that $F$ takes values in the space of integrable vector fields on $\mathbb{R}^3$. The domain and range spaces for $F$ are therefore each Banach spaces with obvious norms. The Fr\'echet derivative of $F$ with respect to $\bm{u}$ is
\begin{align}
D_{\bm{u}}F(\bm{u},\bm{B})[\delta\bm{u}]=\frac{d}{d\epsilon}\bigg|_0(\bm{u}+\epsilon\delta\bm{u})\times\bm{B}=\delta\bm{u}\times\bm{B}.
\end{align}

\section{Simplifying the symplectic part of a phase-space Lagrangian\label{transformation_theory}}
The basic idea that enables a systematic computation of the transformation used in Section \ref{Hamiltonian_theory} is best understood in terms of differential forms. Essentially the same idea is described in the proof of Darboux's theorem in Ref.\,\onlinecite{Abraham_2008}. Let $\Theta=\theta+\delta\theta$ be a $1$-form on a manifold $M$ and suppose that $\omega=-\mathbf{d}\theta$ is non-degenerate. Non-degeneracy means that the mapping $X\mapsto\iota_X\omega$, where $X$ is a vector field on $M$, is injective. Suppose further that $\delta\theta\ll\theta$.  Under these hypotheses, it is possible to find a near-identity transformation $\Phi$ of $M$ that transforms the $2$-form $\Omega=\omega+\delta\omega=-\mathbf{d}\theta-\mathbf{d}\delta\theta$ into the two-form $\omega$. The trick is to express the inverse of transformation as the $\lambda=1$ flow map (here $\lambda$ is being used as the time parameter for the flow map in order to distinguish it from the physical time $t$) associated with a time-dependent vector field $G_\lambda$. Let $F_\lambda$ be the flow map. Now choose $G_\lambda$ in such a way that
\begin{align}\label{lie}
F^*_\lambda(\omega+\lambda \delta\omega)=\omega
\end{align}
for each $\lambda\in[0,1]$. This task may be accomplished if $G_\lambda$ is chosen to be the unique solution of
\begin{align}\label{transform}
\iota_{G_\lambda}(\omega+\lambda\delta\omega)=\delta\theta.
\end{align}
Note that a solution of (\ref{transform}) is guaranteed to exist (at least in finite dimensions) by virtue of three key facts: (1) $\omega$ is non-degenerate, (2) $\lambda\delta\omega\ll\omega$ for $\lambda\in [0,1]$, and (3) the set of non-degenerate two-forms is open in the space of all $2$-forms. Now set $\Phi=F_1^{-1}$. When this transformation is applied to the manifold $M$, the $2$-form $\Omega$ transforms into the $2$-form
\begin{align}
\bar{\Omega}=\Phi_*\Omega=F_1^*\Omega=F_1^*(\omega+\delta\omega)=\omega,
\end{align}
which proves the claim.

When the $2$-form $\Omega$ represents the symplectic structure of a Hamiltonian system on $M$ with Hamiltonian $H$, the transformation $\Phi$ transforms the Poisson bracket into the Poisson bracket associated with the $2$-form $\omega$. Moreover, the Hamiltonian is transformed into
\begin{align}\label{trans_ham}
\bar{H}=\Phi_*H=F_1^*H=H+\int_0^1 L_{G_\lambda}H\,d\lambda+\int_0^1\int_0^\lambda L_{G_{\bar{\lambda}}}L_{G_\lambda}H\,d\bar{\lambda}\,d\lambda+\dots,
\end{align}
where the right-hand-side represents a time-ordered exponential of the vector field $G_t$. Because $G_t$ must be small, Eq.\,(\ref{trans_ham}) provides an asymptotic expansion for the transformed Hamiltonian. 

This general theory reproduces the results of Section \ref{Hamiltonian_theory} when $M$ is taken as the space of tuples $(g_i,g_e,\bm{u},\bm{A})$, with $g_{i,e}$ the ion and electron fluid configuration maps, and the $1$-form $\Theta$ is given by
\begin{align}\label{big_theta}
\Theta[\bm{v}_i,\bm{v}_e,\dot{\bm{u}},\dot{\bm{A}}]=&\int (Z_i n_i +\epsilon^2 \epsilon_o q_e^{-1}\nabla\cdot\bm{E}_\epsilon^*)(m_e \bm{u}_{e\epsilon}^*+\epsilon^{-1}q_e \bm{A})\cdot\bm{v}_e\,d^3\bm{x}\nonumber\\
&+\int n_i (m_i\bm{u}_i-\epsilon^{-1}Z_i q_e \bm{A})\cdot\bm{v}_i\,d^3\bm{x}-\int\epsilon \epsilon_o\bm{E}_\epsilon^*\cdot\dot{\bm{A}}\,d^3\bm{x}.
\end{align}
Here $\dot{\bm{v}}_{i,e}=\dot{g}_{i,e}\circ g_{i,e}^{-1}$. This explains, conceptually at least, why it is possible to find a coordinate transformation of the slow manifold that leads to a truncated expression for the transformed Poisson bracket. 

For the sake of actually performing the computation of a transformation that simplifies $1$-form $\Theta$, instead of applying Eq.\,(\ref{transform}) directly, it is helpful to formulate the problem in terms of iterated transformations as in Ref.\,\onlinecite{Littlejohn_1982}. The remainder of this appendix will explain this computation-oriented approach to find the simplifying transformation. 

Instead of representing the transformation $\Phi$ as the $\lambda=1$ flow of a time-dependent vector field, instead introduce the ansatz
\begin{align}
\Phi=\dots\circ\exp(\epsilon^3 G_3)\circ \exp(\epsilon^2 G_2)\circ\exp(\epsilon G_1),
\end{align}
where $G_1,G_2,\dots$ are time-\emph{independent} vector fields on the space of tuples $(g_i,g_e,\bm{u},\bm{A})$. Such vector fields, which are defined on an infinite-dimensional space, may be represented as a tuple of vector fields on configuration space, $(G_k^i,G_k^e,G_k^{\bm{u}},G_k^{\bm{A}})$, where each entry in the tuple is a functional of $(g_i,g_e,\bm{u},\bm{A})$ that takes values in the space of vector fields on configuration space. In particular, in terms of the component vector fields, the ODE
\begin{align}
\frac{d}{d\lambda}(g_i,g_e,\bm{u},\bm{A})=G_k(g_i,g_e,\bm{u},\bm{A})
\end{align}
may be written
\begin{align}
\frac{d}{d\lambda}g_i&=G_k^{i}\circ g_i\\
\frac{d}{d\lambda}g_e&=G_k^{e}\circ g_e\\
\frac{d}{d\lambda}\bm{u}&=G_k^{\bm{u}}\\
\frac{d}{d\lambda}\bm{A}&=G_k^{\bm{A}}.
\end{align}
Next, expand the $1$-form $\Theta$ in Eq.\,(\ref{big_theta}) in a power series
\begin{align}
\Theta=\epsilon^{-1}\Theta_{-1}+\Theta_0+\epsilon\Theta_1+\epsilon^2\Theta_2+\dots,
\end{align}
and apply the transformation $\Phi$. Modulo unimportant exact differentials, the $1$-form $\Theta$ transforms into
\begin{align}
\bar{\Theta}=\epsilon^{-1}\bar{\Theta}_{-1}+\bar{\Theta}_0+\epsilon \bar{\Theta}_1+\epsilon^2 \bar{\Theta}_2+\dots,
\end{align}
where
\begin{align}
\bar{\Theta}_{-1}&=\Theta_{-1}\\
\bar{\Theta}_{0}&=\Theta_{0}-\iota_{G_1}\mathbf{d}\Theta_{-1}\\
\bar{\Theta}_{1}&=\Theta_{1}-\iota_{G_1}\mathbf{d}\Theta_0-\iota_{G_2}\mathbf{d}\Theta_{-1}\\
\bar{\Theta}_{2}&=\Theta_{2}-\iota_{G_2}\mathbf{d}\Theta_0-\iota_{G_3}\mathbf{d}\Theta_{-1}-\iota_{G_1}\mathbf{d}\Theta_1+\frac{1}{2}(\iota_{G_1}\mathbf{d})^2\Theta_0\\
\bar{\Theta}_{3}&=\dots.
\end{align}
The vector fields $G_k$, which are usually referred to as Lie generators, are finally determined by demanding that the transformed $1$-form truncates at finite-order in $\epsilon$. There is a huge amount of freedom in this step. The prescription for finding the particular transformation that lead to the simple transformed $1$-form found in Eq.\,(\ref{transformed_L}) is given by 
\begin{gather}
\iota_{G_1}\mathbf{d}\Theta_{-1}=0\\
\bar{\Theta}_{k}=0\quad k>0.
\end{gather}
These equations may be solved order-by-order in order to determine each of the $G_k$. The first two Lie generators turn out to be
\begin{align}
G_1^e&=0\\
G_1^i&=0\\
G_1^{\bm{A}}&=0\\
G_1^{\bm{u}}&=\frac{v_A^2}{c^2}\bm{u}_{i\perp}+\frac{\nu Z_i}{1+\nu Z_i}\frac{\mu_o^{-1}\nabla\times\bm{B}}{q_e Z_i n_i},
\end{align}
and
\begin{align}
G_2^e&=-\frac{1}{1+\nu Z_i}\frac{\epsilon_o \bm{u}_i\times\bm{B}}{q_e Z_i n_i}\\
G_2^i&=\frac{\nu Z_i}{1+\nu Z_i}\frac{\epsilon_o \bm{u}_i\times\bm{B}}{q_e Z_i n_i}\\
G_2^{\bm{A}}&=\frac{1}{1+\nu Z_i}\nabla\left(\bm{A}\cdot\frac{\epsilon_o\bm{u}_i\times\bm{B}}{q_e Z_i n_i}\right)\nonumber\\
&-\frac{m_i}{ q_e Z_i}\left((1-\nu Z_i)\frac{v_A^2}{c^2}\bm{u}_i\times\bm{B}-\frac{\nu Z_i}{1+\nu Z_i}\frac{\mu_o^{-1}\nabla\times\bm{B}}{q_e Z_i n_i}\right)\\
G_2^{\bm{u}}&=-\frac{1}{2}\frac{v_A^2}{c^2}G_{1\perp}^{\bm{u}}+\frac{v_A^2}{c^2}\frac{\bm{E}_1^*\times\bm{B}}{|\bm{B}|^2}-\frac{\nu Z_i}{1+\nu Z_i}\left(\frac{\epsilon_o \dot{\bm{E}}_{\text{MHD}}+\rho_{\text{MHD}}\bm{u}_i}{q_e Z_i n_i}\right).
\end{align}

%\bibliography{/Users/josh/Dropbox/Apps/Texpad/latex/cumulative_bib_file.bib}
%\bibliography{cumulative_bib_file.bib}
%%%%%%%%%%%%%%%%%%%%%%%%%%%%%%%%%%%%%

%% put content of bib file here when ready to submit

\providecommand{\noopsort}[1]{}\providecommand{\singleletter}[1]{#1}%
%

%%%%%%%%%%%%%%%%%%%%%%%%%%%%%%%%%%%%

\end{document}